\documentclass{epl}
\usepackage{amssymb,graphicx,psfrag,amsmath}

\title{Boundary effect on CDW: Friedel oscillations, STM image}
\shorttitle{Boundary effect in CDW}
\author{Bal\'azs D\'ora}
\institute{Department of Physics, Budapest University of Technology and Economics,\\
 H-1521 Budapest, Hungary}
\shortauthor{Bal\'azs D\'ora \etal}


\date{}

\begin{document}
\maketitle
\begin{abstract}
We study the effect of open boundary condition on charge density waves (CDW). The electron density 
oscillates 
rapidly close to the boundary, and additional non-oscillating terms ($\sim\ln(r)$)
appear.
The Friedel oscillations survive beyond the CDW coherence length 
($v_F/\Delta$), but their amplitude gets heavily suppressed. The scanning tunneling microscopy image (STM) 
of CDW shows clear features of the boundary. The local tunneling conductance becomes asymmetric with 
respect to the Fermi energy, and considerable amount of spectral weight is transferred to the lower gap 
edge. Also it exhibits additional zeros reflecting the influence of the boundary.

\end{abstract}

\section{Introduction}
The behaviour of charge density waves (CDW) in bulk systems has received considerable attention in the last 
few decades\cite{gruner}. But unlike mesoscopic superconductors, only a few attempts have been made to 
explore the 
response of CDW in mesoscopic systems\cite{dong,gabovich,visscher,tanaka}. These mainly concern charge 
density wave junctions, and concentrate 
on tunneling between CDW and other systems. In this letter, we are going to study the effect of an open end 
along the CDW chain on the electron density and scanning tunneling microscope 
(STM) image, as shown in fig. \ref{stm}. 

As a model we consider a quasi-one dimensional charge density wave, treated in the mean-field 
approximation. Due to quasi-one dimensionality, the effective mass of the electrons in the 
interchain direction is much larger than along the chains, and it is reasonable to assume that
the main affect of neighbouring chains is to suppress the thermal fluctuations of the order 
parameter\cite{gruner,visscher}.
Impurities in low dimensional systems tend to cut the sample. So instead of introducing an 
 external potential representing the impurity, we shall mimic its effect by a semi infinite chain, 
namely by open boundary condition\cite{busch}. Hence the present model is thought to describe impurities in 
the strong pinning limit, or the behaviour of finite CDW with open boundaries. 
Similar models in one dimensional systems have extensively 
been studied over the years\cite{fabrizio,eggert1,eggert}. 

Experimentally, in-chain tunneling studies have been made on CDW systems\cite{slot,zant},
and most of STM scans were performed on bulk CDW materials such as Fe doped
NbSe$_3$\cite{dai} or NbSe$_2$\cite{sacks}, where boundary condition or single
impurity results were not reported. CDW film growth and structuring have been 
started from oxide Rb$_{0.30}$MoO$_3$ (blue bronze)\cite{zant,latyshev} and NbSe$_3$, 
$o$-TaS$_3$\cite{zant1}, but 
no exhaustive studies
of STM imaging were made so far close to interfaces or insulating barriers. Friedel oscillations in 
vanadium-doped blue bronze\cite{rouziere} have successfully been detected around V substituant.

In a normal metal with open boundary (semi infinite chain), the electron density behaves as
\begin{equation}
n(r)=\frac{k_F}{\pi}\left(1-\frac{\sin(2k_Fr)}{2k_Fr}\right),
\label{elso}
\end{equation}
where $r$ is measured from the open end. It vanishes right at the boundary, and produces density 
oscillation with a periodicity of $\pi/k_F$. In the followings we are going to study what happens to a 
CDW close to an open boundary.

\section{Friedel oscillations}

\begin{figure}[h!]
\psfrag{tip}[][b][1][0]{tip }
\psfrag{a}[b][][1.2][0]{A}
\psfrag{ch}[b][][1][0]{chain direction}
\psfrag{cdw}[][][1][0]{CDW sample}
\twofigures[width=6cm,height=6cm]{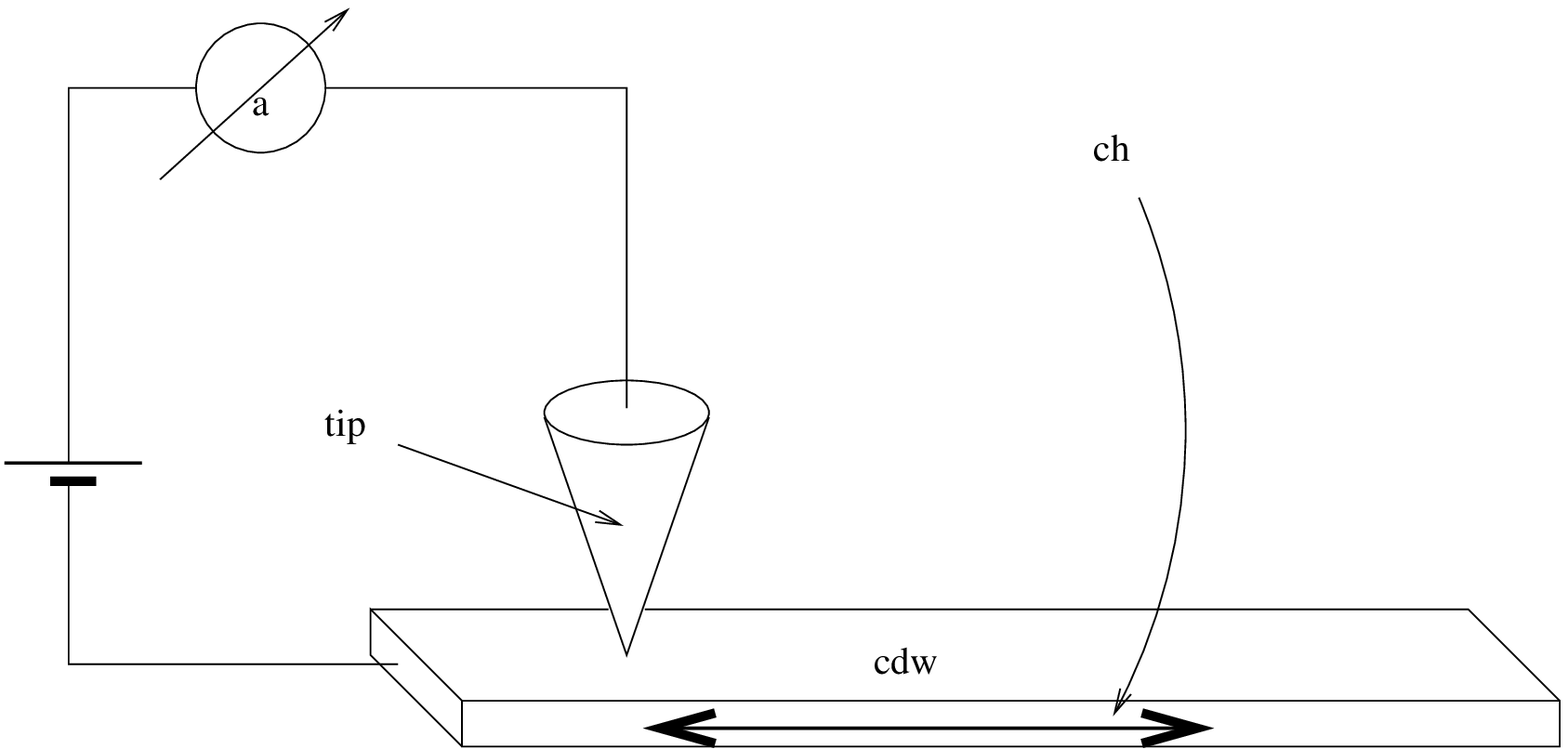}{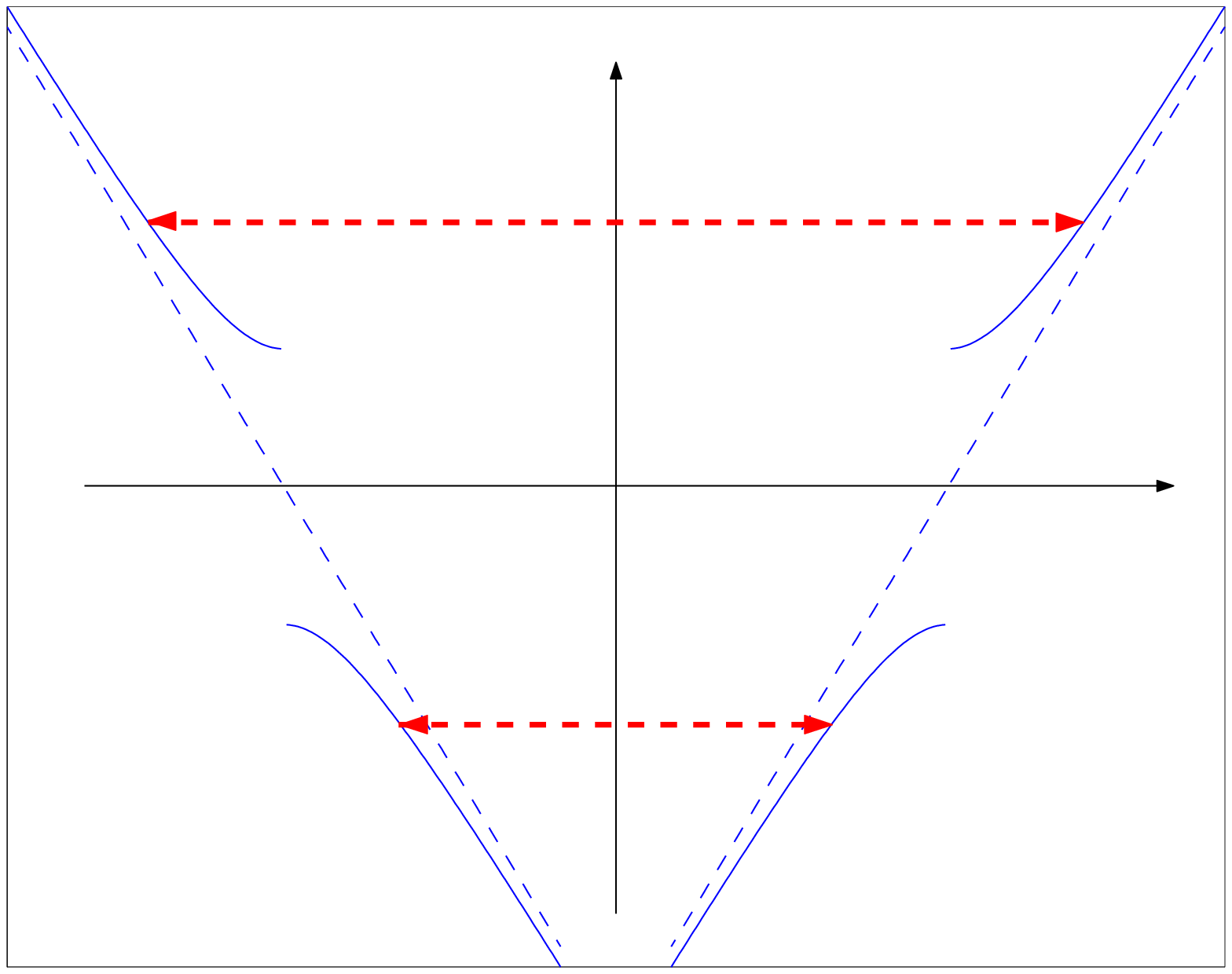}
\caption{Schematic view of the experimental setup. The STM tip moves along the CDW sample.}
\label{stm}
\caption{The schematic energy spectrum of CDW is plotted. Due to the boundary condition, left and right 
movers are connected as shown by the red dashed line.}
\label{spectrum}
\end{figure}

For simplicity we consider a system of spinless fermions in CDW state. The inclusion of spin does not 
alter our results. The effect of open boundary can readily be incorporated into the theory by 
requiring, that the field 
operators should vanish at the origin: $a_{r=0}=0$. This can be fulfilled by connecting the states with 
different wavenumber but equal energy\cite{fabrizio}, as sketched in fig.~\ref{spectrum}. 
Mathematically, this means, that in the reduced Brillouin zone ($0<k<2k_F$), the new quasiparticle 
operators, which diagonalize the CDW system\cite{gruner,nagycikk}, fulfill the relations
\begin{gather}
d_{+,k>k_F}=-d_{+,Q-k}\\
d_{-,k>k_F}=d_{-,Q-k},
\end{gather}
and the original electron operators are expressed as
\begin{eqnarray}
 \begin{array}{c}
       a_{k} \\ a_{k-Q}
 \end{array}
 =\begin{array}{c}
    e^{i\phi}(u_k  d_{+,k}+ v_k  d_{-,k}) \\
    v_k d_{+,k}- u_k d_{-,k}
   \end{array}\label{inverzd}
\end{eqnarray}
with 
\begin{gather}
 \begin{array}{c}
    u_k \\ v_k
 \end{array}
 = \sqrt{\dfrac{1}{2}\left(1 \pm \dfrac{\xi({k})}
 {\sqrt{\xi({k})^2+\Delta^2}}\right)},
\end{gather}
$Q=2k_F$ is the nesting vector, $k_F$ is the Fermi wavenumber, $\xi(k)=v_F(k-k_F)$, $v_F$ is the Fermi 
velocity, $\Delta$ is the CDW 
order parameter.
The phase of the CDW is locked to its optimum value determined by the proximity of the boundary, namely 
$\phi=0$. This assures that the CDW contribution to the electron density at the boundary is minimal.
The very same results can be read off from refs.~\cite{tutto,cheng}, where the effect of point-like 
impurities were explored.
Since we are not interested in phenomenon involving collective 
modes such as the sliding of CDW, only static (electron density) quantity and
tunneling perpendicular to the chain will be considered, the above simplification
is well motivated. 
Due to the proximity of boundary, $\Delta(x)$ is not constant close to the boundary, but decays smoothly 
over a finite distance from the boundary. However, the phenomena discussed here are not expected to be 
altered significantly by position dependent corrections to $\Delta$\cite{visscher,tanaka}. 
  
\begin{figure}[h!]
\psfrag{x}[t][b][1][0]{$2k_Fr$}
\psfrag{y}[b][t][1][0]{$n_{boundary}(r)/n_0$}
\psfrag{t1}[l][r][1][0]{$\Delta/W\ln(2W/\Delta)-1+2(k_Fr)^2/3$}
\psfrag{t2}[][][1][0]{$\Delta/W\ln(v_F/r\Delta)-\sin(2k_Fr)/(2k_Fr)$}
\onefigure[width=9cm,height=9cm]{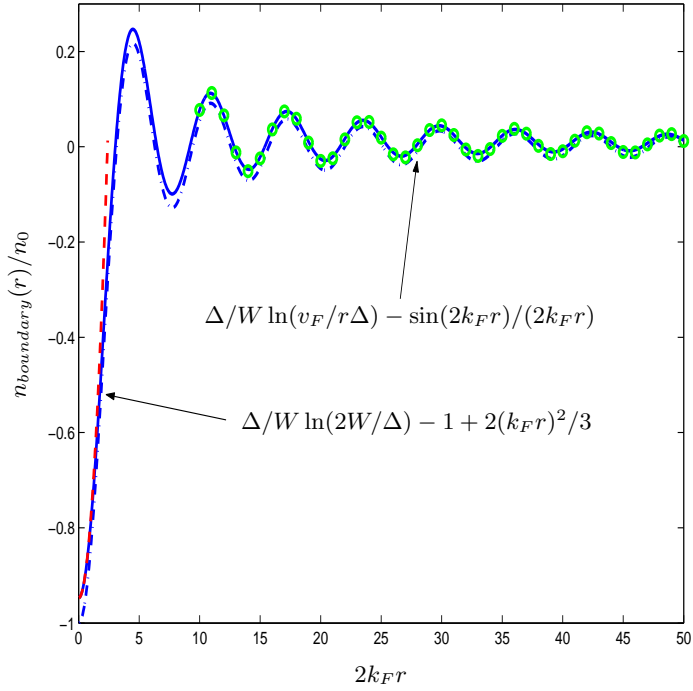}
\caption{The Friedel oscillations caused by the boundary are shown in a CDW for $\Delta/W=0.01$ (solid 
line) and in a normal metal (dashed-dotted line), $n_0=k_F/\pi$ is 
the 
density in a homogeneous system. The red dashed 
line and green circles denote the asymptotic formulas for small and large $r$.}
\label{osc1}
\end{figure}

\begin{figure}[h!]
\psfrag{x}[t][b][1][0]{$2k_Fr$}
\psfrag{y}[b][t][1][0]{$n_{boundary}(r)/n_0$}
\onefigure[width=6cm,height=6cm]{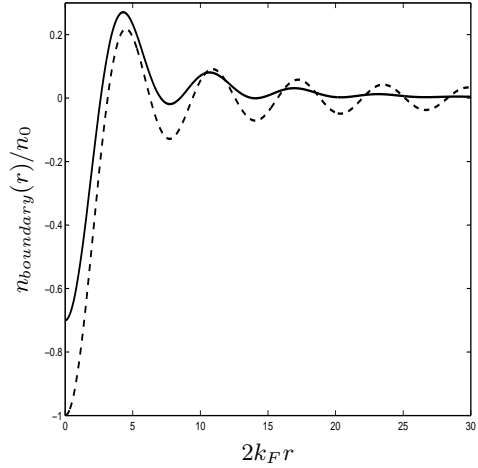}
\caption{The Friedel oscillations caused by the boundary are shown in a CDW for $\Delta/W=0.1$ (solid line) 
and in a normal metal (dashed line). Due to the shorter 
coherence length, the oscillations die out at shorter distances as opposed to fig.~\ref{osc1}.}
\label{osc2}
\end{figure}

The electron density is obtained as 


\begin{gather}
n(r)=2\sum_{0<k<2k_F}\left[\sin^2\left(\frac{\xi(k)r}{v_F}\right)\cos^2(k_Fr)
+\sin^2(k_Fr)\cos\left(\frac{\xi(k)r}{v_F}\right)+\right.\nonumber\\
-\left.\frac{1}{2E}\tanh\frac{\beta 
E}{2}\left(\xi(k)\sin\left(\frac{2\xi(k)r}{v_F}\right)\sin(2k_Fr)
-\Delta\left(\cos(2k_Fr)-\cos\left(\frac{2\xi(k)r}{v_F}\right)\right)\right)\right],
\end{gather}
where $r$ is measured from the open boundary, $E=\sqrt{\xi(k)^2+\Delta^2}$. With the use of the gap 
equation\cite{nagycikk}, this can be further simplified.
At $T=0$, the total density is evaluated as
\begin{gather}
n(r)=\frac{k_F}{\pi}-\frac{2\Delta}{g}\cos(2k_Fr)+\frac{\Delta}{\pi v_F}F(2k_Fr),
\label{suruseg}
\end{gather}
where
\begin{equation}
F(x)=\textmd{K}_0\left(\frac{x\Delta}{W}\right)+\textmd{Ci}(x)-\sin(x)\textmd{K}_1\left(\frac{x\Delta}{W}\right),
\end{equation}
$\textmd{Ci}(x)$ is the cosine integral, $\textmd{K}_n(x)$ is the $n$th Bessel function of the second 
kind, $W=v_Fk_F$, $g>0$ is the effective electron-phonon coupling. The first term in 
eq.~\ref{suruseg} is 
the homogeneous 
electron density, the second one describes the spatially periodic charge density oscillations, while 
the $F(x)$ function contains all the information concerning the effect of the boundary. It is shown 
in figs.~\ref{osc1}, \ref{osc2} for $\Delta/W=0.01$ and $0.1$ together
with the oscillations in a normal metal. $n_{boundary}/n_0=\Delta F(2k_Fr)/W$, 
$n_0=k_F/\pi$ is the density in a homogeneous system. The difference between a normal metal and 
CDW becomes more pronounced for larger values of $\Delta/W$.

Right at the boundary, $n(r=0)=0$ as required. In the $1/k_F\ll r \ll v_F/\Delta$ region, the Friedel 
oscillations caused by the boundary decay as
\begin{equation}
n_{boundary}(x)=\frac{\Delta}{\pi 
v_F}\left(\ln\frac{v_F}{r\Delta}-\gamma\right)-\frac{\sin(2k_Fr)}{2\pi r},
\label{kozep}
\end{equation}
where $\gamma=0.57721$ is the Euler's constant. The first term represents
 the CDW contribution, while the 
second one is identical to that in a normal metal\cite{tutto} as shown in eq.~\ref{elso}. Spatially 
oscillating terms arise also from CDW, but their amplitude are typically $\Delta/W$ times smaller than the 
second term in eq.~\ref{kozep}.

On the other hand, beyond the CDW coherence length,
\begin{equation}
n_{boundary}(r)=\frac{\Delta}{\pi v_F}\frac{\sin(2k_Fr)}{2k_Fr}.
\end{equation}
This means, that the amplitude of the Friedel oscillations changes by a factor of $\Delta/W$ as one passes 
through the CDW coherence length from the boundary. This makes the detection of such fine structures 
superimposed on the usual charge density oscillations very difficult. On the other hand, the short distance 
behaviour could readily be checked experimentally. A more direct way to detect boundary effects is to look 
for probes with energy and spatial resolution such as the local tunneling conductance, as will be 
discussed in the followings.

\section{STM image, tunneling conductance}

\begin{figure}[h!]
\psfrag{x}[t][b][1][0]{$V/\Delta$}
\psfrag{y}[b][t][1][0]{$k_Fr$}
\onefigure[width=9cm,height=9cm]{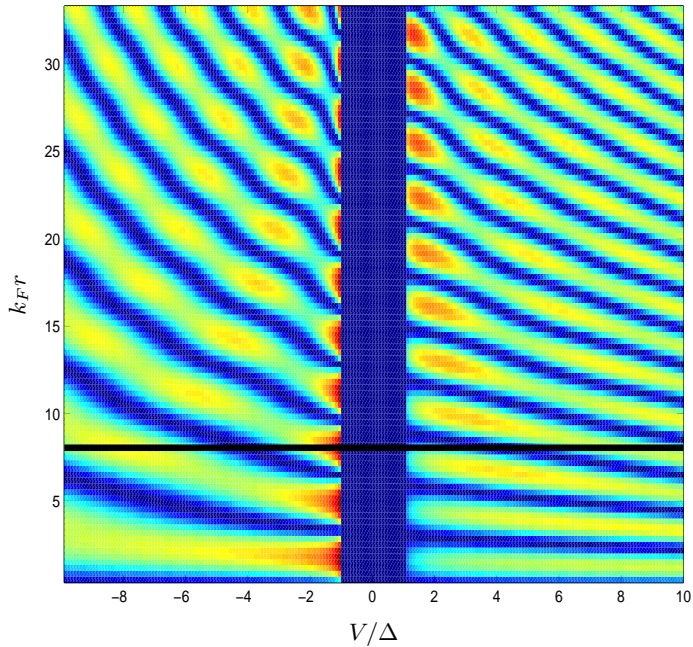}
\caption{The local tunneling conductance is shown close to the boundary as a function of voltage ($V$)
and distance from the boundary ($r$) for $W/\Delta=20$. The tunneling current along the horizontal black
line is plotted in fig.~\ref{metszet}.}
\label{ldosneg}
\end{figure}

\begin{figure}[h!]
\centering
\psfrag{x}[t][b][1][0]{$V/\Delta$}
\psfrag{yy}[b][t][1][0]{$dI(r)/dV$ (arb. units)}
\psfrag{y}[b][t][1][0]{$I(r)$ (arb. units)}
\includegraphics[width=90mm,height=90mm]{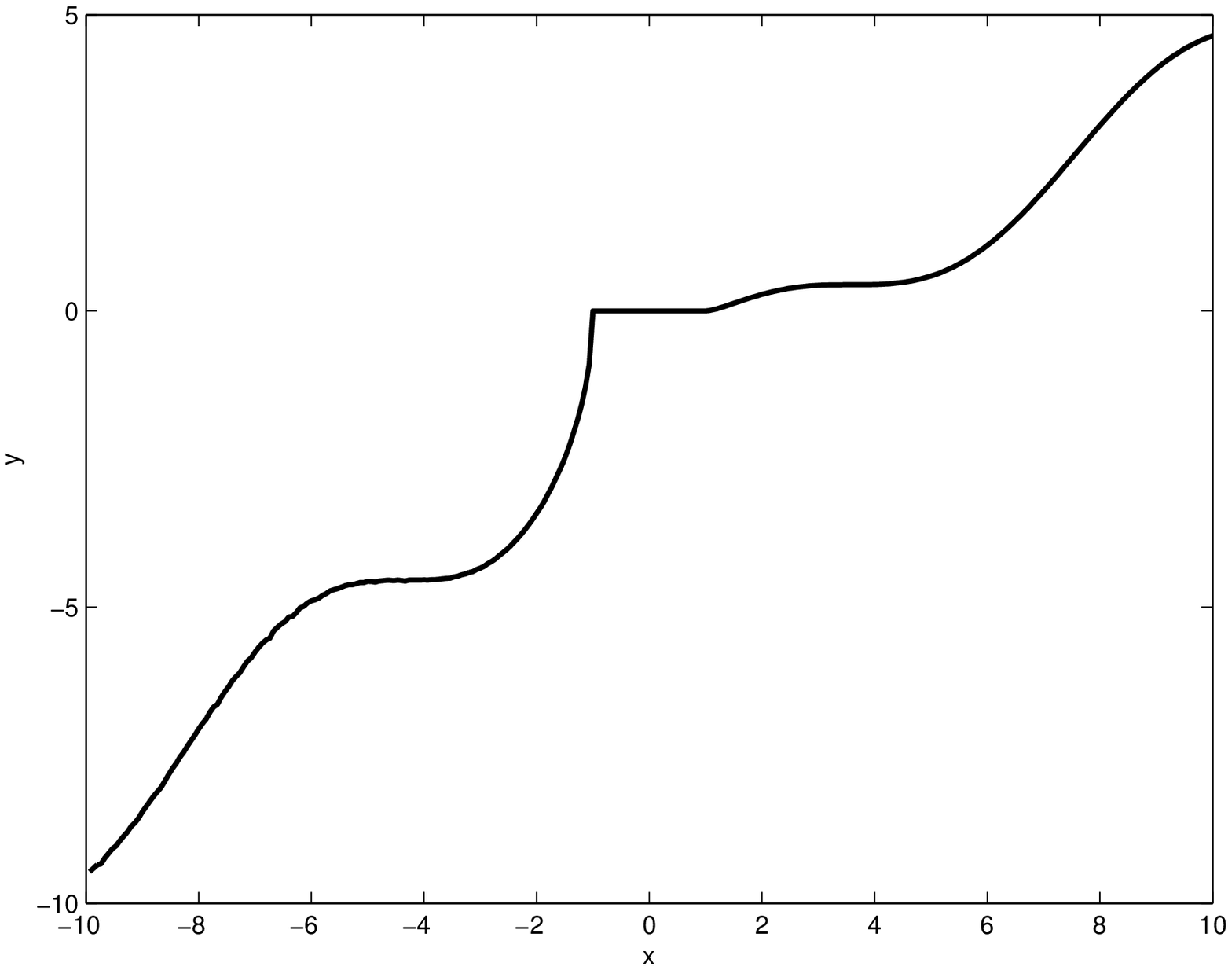}

\vspace*{-54mm}\hspace*{43mm}\includegraphics[width=42mm,height=42mm]{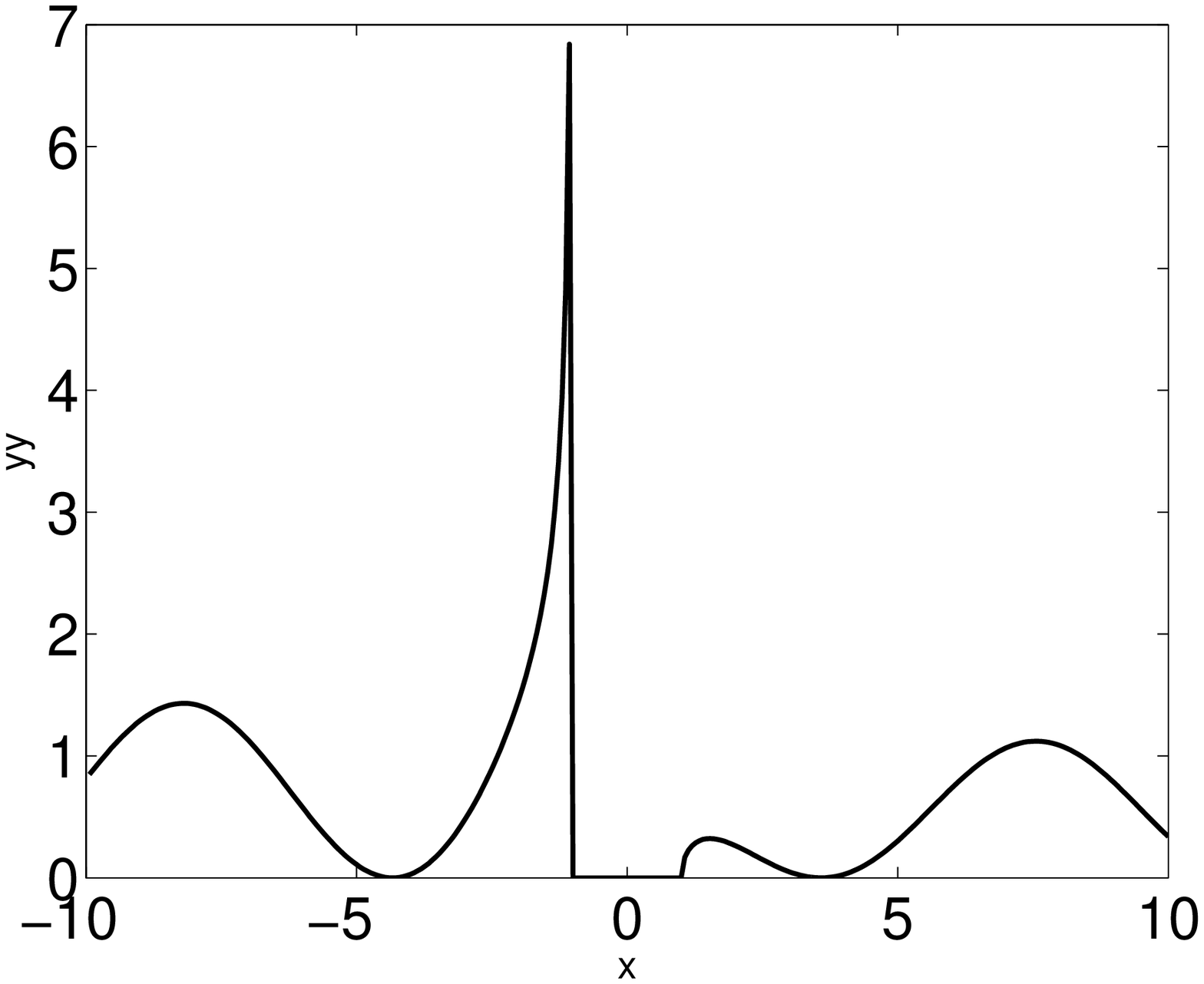}

\vspace*{15mm}

\caption{The STM tunneling current ($I(r)$) is shown at $k_Fr=8$ as a function of the applied voltage along 
the 
black line in fig.~\ref{ldosneg}. The inset shows the tunneling conductance ($dI(r)/dV$) along the 
same line.}
\label{metszet}
\end{figure}

The 
STM current $I$ is directly related to how many electrons are locally available in the CDW.
 At a position $r$, the local tunneling conductance measures the local density of states, 
and is given as
\begin{equation}
\frac{dI(r)}{dV}\sim N(\omega,r)=\frac{1}{2\pi}\int\limits_{-\infty}^\infty e^{i\omega 
t}\langle\left\{a_r(t),a^+_r(0)\right\}\rangle dt.
\label{differ}
\end{equation}
Usually the tunneling matrix 
element between the 
sample and the tip depends on the wavevectors of the electron in the CDW and 
in the tip. However, the 
behaviour of CDW is mainly determined by electrons living in the Fermi 
surface, 
which is determined by the wavevector component corresponding to the quasi-one 
dimensional direction. In this sense, this component carries all the 
informations about the condensate, and the inclusion of perpendicular 
components is not expected to alter the results.
After straightforward calculation eq.~\ref{differ} yields to
\begin{gather}
\frac{dI(r)}{dV}\sim
\frac{4|\omega|}{\pi 
\sqrt{\omega^2-\Delta^2}}\left(\sin^2(\sqrt{\omega^2-\Delta^2}r/v_F)\cos^2(k_Fr)
+\sin^2(k_Fr)\cos^2(\sqrt{\omega^2-\Delta^2}r/v_F)+\right.
\nonumber\\
+\left.\frac{\sqrt{\omega^2-\Delta^2}}{2\omega}\sin(2\sqrt{\omega^2-\Delta^2}r/v_F)
\sin(2k_Fr)+\frac{\Delta}{2\omega}(\cos(2k_Fr)-\cos(2\sqrt{\omega^2-\Delta^2}r/v_F))\right),
\end{gather}
with $E=\sqrt{\omega^2-\Delta^2}$. In a normal metal this reduces to
\begin{equation}
\frac{dI(r)}{dV}\sim \frac 4\pi\sin^2\left(\frac{\omega r }{v_F}+k_Fr\right).
\end{equation}
Hence the position of zeros in both cases is determined as $r=v_F n\pi/(\omega+W)$, $n$ a natural number. 
As 
$\omega$ increases, the pattern gets denser, as can be checked in fig.~\ref{ldosneg}. Along the black
 line, a typical plot of the local density of states and the tunneling current is shown in 
fig.~\ref{metszet}. The tunneling current remains unchanged with varying voltages at the zeros of the local 
density of states, namely at $\omega=\dfrac{v_Fn\pi}{r}-W$.
In a homogeneous CDW, we expect sharp peaks at both $\Delta$ and $-\Delta$. In the present case, however, 
 due to the presence of bound states caused by the boundary, only one peak remains situated at 
$-\Delta$, in accordance with ref. \cite{cheng}. Here the effect of a single impurity was studied. After 
letting the strength of impurity to go to infinity, the bound state induced by the scatterer moves to 
$-\Delta$, and the divergent peak at $\Delta$ disappears.

\section{Conclusion}

We have investigated the effect of an open boundary on charge density waves. It is believed that 
the effect of strong impurities or tunneling barriers can be approximated by an open end.
The right and left moving particles mix up. The electron density consists of the usual CDW contribution and 
terms caused by the proximity of boundary. Within the CDW coherence length ($\sim v_F/\Delta$), these terms 
correspond to the 
response of a free electron gas and additional $\ln(r)$ corrections caused by the interaction between the 
boundary and the condensate. Beyond the CDW coherence length, only the small amplitude oscillations 
survive, but their detection seems to be a very difficult task to deal with.
Quantities with both energy and spatial resolution might better help to clearly see boundary effects.
Among them we have chosen to study the local tunneling conductance along the chain, measurable by scanning 
tunneling microscope. The differential conductance is zero for voltages smaller than the gap maximum. For 
negative $V$, $dI(r)/dV$ exhibits a sharp peak $\sim \Delta/\sqrt{V^2-\Delta^2}$ as $V$ approaches 
$-\Delta$. For positive voltages, the differential conductance increases smoothly with $V$, and no 
divergences are found. This is in accordance with ref. \cite{cheng}. The tunneling current changes 
inflection with increasing tip voltage due to the presence of zeros in the local density of states of CDW. 
Also it remains unchanged not only around the Fermi energy, but also at the additional zeros of the local 
density of states given by $\omega=\dfrac{v_Fn\pi}{r}-W$.

We are indebted to Istv\'an Nagy, Andr\'as V\'anyolos and Attila Virosztek for
useful discussions. This work was supported by the Magyary Zolt\'an postdoctoral
program of Foundation for Hungarian Higher Education and Research (AMFK) and by
the Hungarian
Scientific Research Fund under grant numbers OTKA TS040878 and TS049881.

\bibliographystyle{apsrev}
\bibliography{eth}

\end{document}